\documentclass[prc,preprint,showpacs,nofootinbib]{revtex4}
\usepackage{graphicx}
\def\para{\uparrow\uparrow,\downarrow\downarrow}
\def\antipara{\uparrow\downarrow,\downarrow\uparrow}
\def\be{-E/A}
\def\bge{\begin{equation}}
\def\ene{\end{equation}}
\def\bg{\begin{eqnarray}}
\def\en{\end{eqnarray}}

\begin{document}

\title{Simulation of symmetric nuclei 
and the role of Pauli potential in binding energies and radii}

\author{M. \'{A}ngeles P\'{e}rez-Garc\'{i}a~$^{1}$~\footnote{mperezga@usal.es}, 
K. Tsushima~$^{2}$~\footnote{tsushima@jlab.org}, 
A. Valcarce~$^{1}$~\footnote{valcarce@usal.es}}

\affiliation{$^{1}$Departamento de F\'{i}sica Fundamental and 
Instituto Universitario de
F\'{i}sica Fundamental y Matem\'{a}ticas, \\IUFFyM, Universidad de Salamanca, 
Plaza de la Merced s/n 37008 Salamanca\\ $^{2}$Thomas Jefferson National Accelerator Facility
12000 Jefferson Avenue, Newport News, VA 23606,USA}

\date{\today}% Deleting this command produces today's date.

\begin{abstract}
It is shown that the use of a density
dependent effective Pauli potential together with a nucleon-nucleon interaction potential plays a crucial role to reproduce not only the
binding energies but also the matter root mean square radii of medium mass range spin-isospin saturated nuclei.
This study is performed with a semiclassical Monte Carlo many-body simulation within the context of a simplified nucleon-nucleon interaction to focus on the effect of the genuine correlations due to the fermionic nature of nucleons. The procedure obtained is rather robust and it does not depend on the detailed features of the nucleon-nucleon interaction. For nuclei below saturation the density dependence may be represented in terms either of the nucleon number, $A$, or the associated Fermi momenta. When testing the simulation procedure for idealized "infinite" symmetric nuclear matter within the corresponding range of densities, it turns out that finite size effects affect the Pauli potential strength parametrization in systems up to about $120$ particles while remaining approximately stable for larger systems.
\end{abstract}
\vspace{1pc}

\maketitle

\section{Introduction}
\label{intro}

We have recently demonstrated~\cite{future} that one can always expect to be able to reproduce  the empirical binding energies for a set of nuclei by introducing a proper density dependent Pauli potential in terms of a single variable, the nucleon number $A$. Such a result is the consequence of a delicate counterbalance between kinetic and potential energetic contributions.  A deeper understanding of the nuclear saturation mechanism and reproduction of other observables for nuclei has been made possible due to the richness of available nuclear data. Since the pioneering shell model calculations~\cite{shell} and the so-called "Coester band"~\cite{Coester}, many microscopic investigations have been devoted to study 
the properties of many-nucleon systems with a variety of approaches as {\it i.e.} relativistic mean fields~\cite{QHD}, quark-based models~\cite{QMC,QCM} and chiral dynamics~\cite{chiral}. On the other hand, studies based on computer simulations 
with Monte Carlo techniques and/or (quantum) molecular dynamics, 
have proved themselves to be very powerful to investigate
many-nucleon systems such as 
subthreshold kaon production~\cite{Aichelin,Fuc06} 
and multifragmentation~\cite{Aichelin,fragment}  
in heavy ion collisions, and 
the {\it nuclear pasta} phase~\cite{pasta,Watanabe,Maruyama,hor}. 
Most of these works, based on semiclassical simulations, treat nuclei 
as composite objects interacting through effective potentials containing 
the essential features of the system. Especially, the effects of 
the Pauli principle, the basic quantum feature for many-fermion 
systems, is simulated by means of an effective potential 
depending on the position and momentum of the interacting nucleons.  
This was firstly suggested by Wilets {\it et al.}~\cite{Wilets} 
in the context of classical many-body nuclear models.
Later, approaches such as fermionic molecular dynamics (FMD)~\cite{FMD} and antisymmetrized 
molecular dynamics (AMD)~\cite{AMD}, include the Pauli 
principle in a quantum mechanical manner. However, it is still not practical 
to apply these methods to 
study heavy nuclei and nuclear matter, due to the need of heavy 
computational calculation, and unestablished (anti)periodic 
boundary conditions to simulate an infinitely large 
nuclear system~\cite{tomo}.

In this work we go a step further from our previous study~\cite{future} by considering a refined effective spin-dependent NN interaction and analyzing other nuclear properties for medium mass nuclei and heavy nuclear systems. We will firstly deduce a corresponding density dependent effective Pauli potential reproducing the empirical nuclear binding energies. For this purpose, without loss of generality, we will focus on medium mass range nuclei with $8 \le A \le 44$. It turns out that the spin dependence has only a very moderate impact, with the relevant contributions coming from the in-medium effects of the effective Pauli potential. In addition, we will also evaluate the root mean square radius (rms) of spin-isospin saturated nuclei and their density distributions. When the simulation is extended for idealized infinite symmetric nuclear matter, it turns out that finite size affects the Pauli potential strength up to about 120 particles while remaining almost stable for larger systems. 
The picture arising from the present semiclassical simulation of nuclei, though simplified, accounts for basic properties of these systems. In the medium mass range nuclei, the density dependence could be either described in terms of the nucleon number or, as we will show, a function of the associated nucleus Fermi momentum. This treatment relates to that of  recent approaches applied in chiral dynamics~\cite{chiral}. The present description can be considered, then, as an improvement of  previous simulation work where neither antisymmetrization effects nor density dependence were considered. The derived effective Pauli potential does not depend on local densities as occurs, for example, in work using Skyrme models ~\cite{Aichelin}.  Nonetheless, even if not local, the density dependence in the effective Pauli potential becomes crucial to achieve saturation. 

The structure of the paper is as follows.
In section~\ref{model}, we describe the different nuclear 
potentials used in the simulations.
First, we introduce $S$-wave NN interactions without spin 
nor isospin dependence. 
This makes easier to analyze the role of the Pauli potential 
and its nuclear density dependence. 
Next, we consider improved spin dependent NN interactions. 
Results will be presented in section~\ref{results}, 
and summary and conclusions will be given in section~\ref{summary}.

\section{Nuclear interaction models}
\label{model}
In this section we present the different simplified 
nuclear models which will be used in our simulations.
We restrict ourselves to spin-isospin saturated, $Z=N$ (even $Z$ and $N$) nuclei 
with mass number $A$ in the range $8 \le A \le 44$.
In addition, nucleons are treated as classical, structureless particles.  

\subsection{Square-well potential}
\label{SWmodel}
First, we consider a NN interaction based 
on a square-well potential without any dependence on spin nor isospin. 
While the model is simple, it retains the basic features of 
the NN interaction, short-range repulsion and intermediate range 
attraction. The NN interaction potential between the $i$-th and $j$-th 
nucleons is given by, 
\begin{equation}
V_{NN}(r_{ij})=
\left\{
\begin{array}{lll}
V_{Core}, &\quad {\rm for} \quad 0 \le r_{ij} < a,  
&\quad \\
- V_0,     &\quad {\rm for} \quad a \le r_{ij} < a+b,
&\quad \\
0,        &\quad {\rm for} \quad a+b \le r_{ij}, &
\end{array}\right. 
\label{swpot}
\end{equation}
where $r_{ij}=|{\bf r}_i-{\bf r}_j|$ is their relative distance.  
This potential consists of a repulsive core of strength $V_{Core}$ and width 
$a$, and an attractive well of depth $V_0$ and width $b$. 
We will consider values $V_{Core}=100$ MeV, $V_0=3$ MeV, $a=2$ fm and $b=2$ fm. The core strength gives only a very moderate contribution to the binding energies (as easily deduced from the different values we may use~\cite{future}), only preventing nucleons from occupying the inner region in configuration space. However the spatial parameters,  $a$ and $b$, are crucial to reproduce satisfactorily matter root mean square radii as will be explained later in this section. The corresponding Hamiltonian is given by, 
\begin{equation}
{H}=\sum_{i=1}^{A} \frac{{\bf p}_{i}^{2}}{2m_N}
+\sum_{i=1,j>i}^{A} V_{NN}(r_{ij}) \, ,
\label{SW}
\end{equation}
where ${\bf p}_{i}$ is the 3-momentum of $i$-th nucleon with mass $m_N$. 
From now on we will refer to this model as SW.

Next, we consider a refined NN interaction including 
the Coulomb potential: 
\begin{equation}
V_{Coul}(r_{ij})=  
\frac{e^2}{4 \pi r_{ij}} (1/2+\tau_i)(1/2+\tau_j),   
\label{Coulombpot}
\end{equation}
where $\tau_i$ ($\tau_j$) is the isospin third-component of
$i$-th ($j$-th) nucleon ($+1/2$ for protons, $-1/2$ for neutrons),  
and $e$ is the proton electric charge. 
Then, the Hamiltonian is given by, 
\begin{equation}
H=\sum_{i=1}^{A} \frac{{\bf p}_{i}^{2}}{2m_N} 
+\sum_{i=1,j>i}^A 
\left[ {V}_{NN}(r_{ij}) + V_{Coul}(r_{ij})\right]. 
\label{SWC}
\end{equation}
We will refer to this model as SWC.

Since nucleons are fermions, they obey the Pauli principle. 
In the present treatment this is mimicked by introducing 
an effective potential, which prevents nucleons 
from occupying the same phase space volume when they have the same 
quantum numbers. This was first suggested 
by Wilets {\it et al.}~\cite{Wilets}.
Later, a Gaussian form of the Pauli potential 
has been introduced by Dorso {\it et al.}~\cite{dorso}.
In this way one can reproduce well 
kinetic energies per nucleon in a Fermi gas, although it 
fails to describe other basic features such as two-body correlation functions 
at low temperature~\cite{fai}. 
Using this effective Pauli potential depending 
not only on the position but also on the momentum of the 
interacting fermions in a Hamiltonian approach, an 
effective nucleon mass which grows largely 
with increasing nucleon density arises.
This results in very slow nucleons, and may lead to quasi-crystallized
nuclear Fermi gases if not properly considered~\cite{newpauli}. 

In this work we use the Pauli potential form proposed by Dorso 
{\it et al.}~\cite{dorso},

\begin{equation}
{V}_{Pauli}({r}_{ij}, {p}_{ij})= 
V_P\,\,  \exp \left(-\frac{r_{ij}^2}{2q_0^2}
-\frac{p_{ij}^2}{2p_0^2}\right) \delta_{\tau_i \tau_j}
\delta_{\sigma_i \sigma_j} \, ,
\label{Paulipot}
\end{equation}
where $p_{ij}=|{\bf p}_i-{\bf p}_j|$ is the relative 3-momentum of 
the $i$-th and $j$-th nucleons, with 
$\delta_{\tau_i \tau_j}$ ($\delta_{\sigma_i \sigma_j}$) being  
the Kronecker's delta for the isospin (spin) third-component. 
We will differentiate two cases:  (i) $V_P$, $q_0$ and $p_0$ 
are constant fixed to reproduce the empirical 
binding energy per nucleon ($\be$) of $^{16}$O, 
7.98 MeV~\cite{nucdata}, 
and (ii) allow them to be density dependent.
The models corresponding to (i) and (ii), will be referred to as 
SWCPo and SWCP, respectively. In addition, in both cases, spatial parameters in Eq.~(\ref{swpot}) take the values $a=2$ fm and $b=2$ fm as they have been adjusted to reproduce the $^{16}$O matter rms radius to $2.72$ fm~\cite{nucdata2}. The Hamiltonian for these models, without specifying the 
density dependence for the Pauli potential, is given by:
\begin{equation}
H=\sum_{i=1}^{A} \frac{{\bf p}_{i}^{2}}{2m_N}
+ \sum_{i=1,j>i}^{A} 
\left[ {V}_{NN}(r_{ij})+V(r_{ij})_{Coul}
+V_{Pauli} (r_{ij},p_{ij}) \right]. 
\label{SWCP}
\end{equation}

To extract the density (Fermi momentum) dependence of
$q_0$ and $p_0$ in the Pauli potential in Eq.~(\ref{Paulipot}), we use the Fermi 
momenta deduced in Ref.~\cite{moniz} from the quasielastic 
electron scattering on several nuclei, and interpolate 
for the nuclei studied later. Note that with this parametrization of the Pauli potential the more general local density dependence has been parametrized in terms of one single parameter based on a simplified Fermi gas picture.

In a nucleus, a typical nucleon sphere radius $r$ may be given by,
\begin{equation}
r=\left(\frac{3}{4 \pi \rho}\right)^{1/3},
\label{distance}
\end{equation}
where $\rho=2 p_F^3/3 \pi^2$ is the nucleon density and 
$p_F$ the Fermi momentum of nucleons in the nucleus. 
Then, the averaged inter-nucleon distance $2r$ 
may be estimated as $(2r/\sqrt{2}q_0)\simeq 1$, 
where $q_0$ is an "effective range" of the Pauli potential. 
This leads to, 
\bg 
q_0&\simeq&\frac{\hbar(9 \pi)^{1/3}}{\sqrt{2}p_F}, 
\label{q0}\\
p_0&\simeq&\frac{\hbar}{q_0} 
=\frac{\sqrt{2}}{(9 \pi)^{1/3}}p_F, 
\label{p0}
\en
with the uncertainty principle for $q_0$ and $p_0$ 
to satisfy $q_0\,\, p_0 \simeq \hbar$. 

As mentioned above, we will first consider
a model where the Pauli potential parameters,
$V_P$, $q_0$, and $p_0$ are
fixed to reproduce the binding energy of $^{16}$O, what
gives rise to ($V_P,q_0,p_0$)=($41$ MeV, $1.88$ fm, $104.96$ MeV/c). 
Secondly, we consider another parametrization of the Pauli potential
allowing a density ($p_F$) dependence trying to reproduce the empirical 
binding energies for 
spin-isospin saturated nuclei, together with the relations in 
Eqs.~(\ref{q0}) and~(\ref{p0}). This leads to,
\bge
V_{Pauli}(r_{ij},p_{ij},p_F)
=V_P(p_F)\,\exp \left[-\frac{r_{ij}^2}{2q_0^2(p_F)}
-\frac{p_{ij}^2}{2p_0^2(p_F)}\right] 
\delta_{\tau_i \tau_j} \delta_{\sigma_i \sigma_j} \,\, {\rm (MeV)} \, .
\label{VPden}
\ene
It is worth noting that this approach has some similarities to what is used in other contexts, such as chiral dynamics~\cite{chiral}.  This may imply that the variable $p_F$ in Eq.~(\ref{VPden}), is appropriate to characterize the many-nucleon system.

\subsection{Spin dependent square-well potential}
\label{SDmodel}

The NN interaction potentials we have described so far  
may be more generally described by introducing spin dependence. Note that this also implies isospin dependence
in order to satisfy the Pauli principle.
We consider a simple potential for the 
spin parallel ($\para$) and antiparallel ($\antipara$) pairs of nucleons: 
\begin{equation}
V^{spins}_{NN}(r_{ij})=
\left\{
\begin{array}{ll}
V_{Core}, &\quad {\rm for} \quad 0 \le r_{ij} < a,\\
- V^{\para}_0\delta_{\sigma_i \sigma_j} 
- V^{\antipara}_0(1-\delta_{\sigma_i \sigma_j}), 
&\quad {\rm for} \quad a \le r_{ij} < a+b,\\
0,        &\quad {\rm for} \quad a+b \le r_{ij}.
\end{array}\right.
\label{vspin}
\end{equation}
We distinguish three different possibilities
for the strength of each spin dependent term:
(i) Model SD1, where the spin parallel potential   
is more attractive than the antiparallel, 
$|V_0^{\para}|>|V_0^{\antipara}|$;  
(ii) Model SD2, the opposite case, $|V_0^{\para}|<|V_0^{\antipara}|$; 
(iii) Model SD3, for the case, $|V_0^{\para}|=|V_0^{\antipara}|$. 
We will consider values $a=1$ fm and $b=2$ fm, different from those used for the models described in Sect.~\ref{SWmodel}. In this way we introduce some dependence on the spatial parameter value $a$ in Eq.~(\ref{swpot}). We have verified that parameter $a$ (and also $b$) is mainly responsible for reproducing rms radii but has a very moderate impact on binding energies. We will mainly focus on the effect of the spin dependence on binding energies.
The different balance between the parallel and antiparallel
strengths will allow us to overcome the possible consequences
of the simplification including only $S$-wave interactions.
Inclusion of also the $D$-wave interactions, which only couple to the
$^3S_1$ channel (in the present case, e.g., the interaction 
between $(p\uparrow)-(n\uparrow)$ pair) would modify the balance
between the attraction in the spin parallel and antiparallel cases. 

Finally, let us mention that although we have not considered a
possible radial dependence on the interactions, in the case of binding energies 
will not have observable effects.
The strength of $V_0^{\para}$ and $V_0^{\antipara}$ for   
the SD1, SD2 and SD3 models are chosen rather arbitrarily 
but they all reproduce the empirical value, $\be=7.98$ MeV 
of $^{16}$O, together with the same Pauli potential in the 
SWCPo model. The Hamiltonian for these models is given by:
\begin{equation}
H= \sum_{i=1}^{A} \frac{{\bf p}_{i}^{2}}{2m_N}
+ \sum_{i=1,j>i}^{A} \left[ {V}^{spins}_{NN}(r_{ij}) +
{V}_{Pauli}({r}_{ij}, {p}_{ij}) + V_{Coul}(r_{ij}) \right] \, .
\label{Hspin}
\end{equation}
In Table~\ref{tab:SDparam} we summarize the
strength of $V_0^{\para}$ and $V_0^{\antipara}$ for 
all models.
\begin{table}[htbp]
\begin{center}
\caption{Strength (in MeV) of $V_0^{\para}$ and $V_0^{\antipara}$ for 
the SD1, SD2 and SD3 models.}
\label{tab:SDparam}
\begin{tabular}{l|lll} \hline
Model &$V_{Core}$   &$V_0^{\para}$ & $V_0^{\antipara}$ \\ \hline
SD1 &100  &5.8 &0.5 \\
SD2 &100  &0.5 &5   \\
SD3 &100  &3   &3 
\end{tabular}
\end{center}
\end{table}

\section{Simulation results}
\label{results}

In this section we present the results obtained using Monte Carlo simulations of the nuclear systems where a typical value for low temperature, $T=1$ MeV, is adopted.
Initially, the nucleons are uniformly distributed  
inside a sphere of radius $R_0$, within a cubic box of  volume $V=L^3$ 
and impose $L >> r_{ij}$. Up to $R_0 \simeq 5$ fm the binding energy per nucleon
has rather stable values, although it tends to vanish rapidly for larger $R_0$. This 
reflects the fact that for large values of $R_0$ some nucleons would not feel
any attraction (see Eq. (\ref{swpot})). 
Then, using the Metropolis algorithm~\cite{Metropolis}, 
the initial random seed is iterated until the energetic 
configuration achieves the lowest possible value. In addition, we must 
check that a clustered configuration of nucleons is indeed formed, a nucleus. 
Once the system is thermalized after performing a sufficiently large  
number (several thousands) of Monte Carlo sweeps, 
we take data to calculate the statistical 
average for thermodynamic quantities such as kinetic and potential energies and  root mean square radii.

First, we discuss the results of the SWCPo model 
introduced in Sect.~\ref{SWmodel}, based on the density independent Pauli potential fixed to reproduce the  binding energy per nucleon and matter rms radius of $^{16}$O. 

\begin{figure}[hbtp]
\begin{center}
\includegraphics [angle=-90,scale=1] {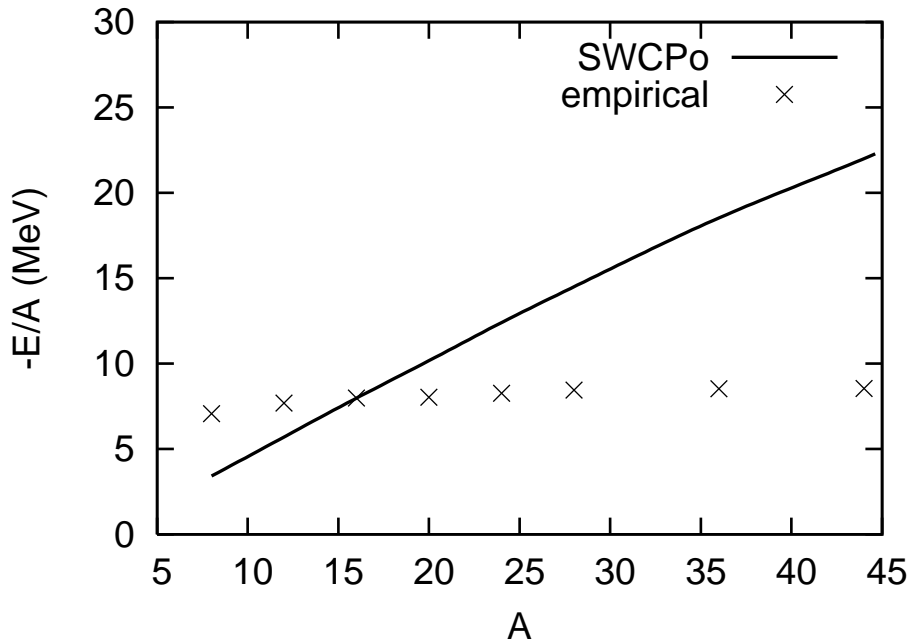}
\caption{Binding energy per nucleon, $\be$, calculated in 
the SWCPo model. Empirical data, shown by crosses, 
are taken from Ref.~\cite{nucdata} for even nuclei. 
We have calculated nuclei with $A=8,12,16, ..., 44$ and 
the solid line goes through all calculated points,
being only valid for these even nuclei. This will also
apply to the other figures along the text.} 
\label{Fig1}
\end{center}
\end{figure}

In Fig.~\ref{Fig1} we show the SWCPo results for $\be$   
by the solid line compared to empirical values~\cite{nucdata} 
(crosses).
%%%%%%%%%%%%%%%%%%%%%%%%%%%%%%%%%%%%%%%%%%%%%%%%%%%%%%%%%%%%%%%%%
%%%%%%%%%%%%%%%%%%%%%%%%%%%%%%%%%%%%%%%%%%%%%%%%%%%%%%%%%%%%%%%%%

As can be seen, using the density independent Pauli potential,
$\be$ grows linearly as the nucleon number increases.
Thus, the empirically observed saturation for $\be$ cannot be achieved.

\begin{figure}[hbtp]
\begin{center}
\includegraphics [angle=-90,scale=0.75] {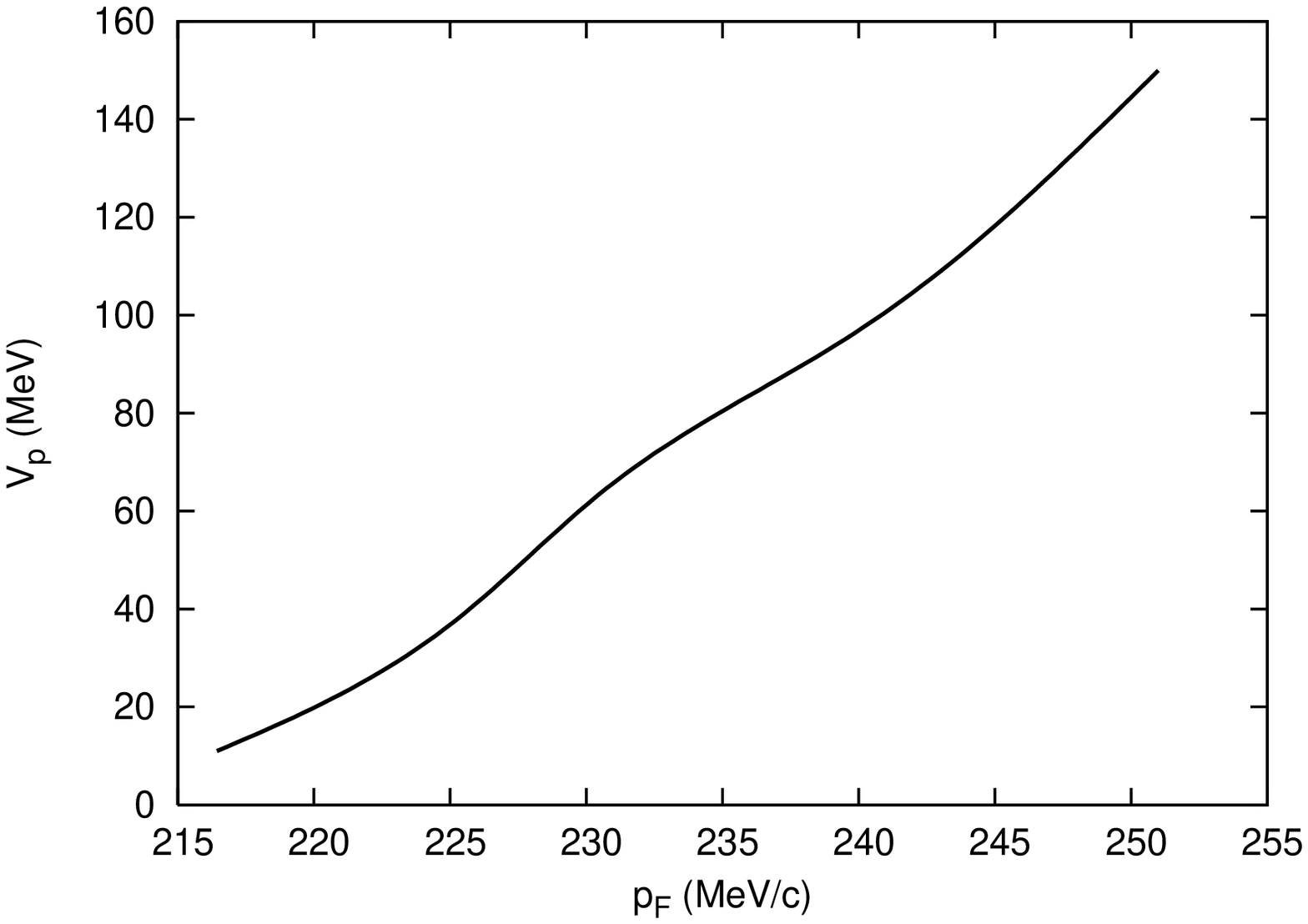}
\caption{Pauli potential strength $V_P$ as a function of the Fermi momentum, $p_F$.}
\label{Fig2}
\end{center}
\end{figure}

Let us now discuss the results for the SWCP model, 
with the density dependent Pauli potential.
Before presenting the results for $\be$, 
we show in Fig.~\ref{Fig2} the $p_F$ for 
$V_P$ obtained to reproduce the empirical $\be$ values for 
the spin-isospin saturated nuclei.
%%%%%%%%%%%%%%%%%%%%%%%%%%%%%%%%%%%%%%%%%%%%%%%%%%%%%%%%%%%%%%%%%
%%%%%%%%%%%%%%%%%%%%%%%%%%%%%%%%%%%%%%%%%%%%%%%%%%%%%%%%%%%%%%%%%
The strength $V_P(p_F)$ increases as Fermi momentum (density) 
increases. The strength obtained for each nuclear species can be mapped as a 
function of the Fermi momentum extracted from the quasielastic
electron scattering off nuclei~\cite{moniz}. This behavior is analogous to 
the vector mean field in Hartree approximation 
in relativistic mean field models~\cite{QHD}.

\begin{figure}[hbtp]
\begin{center}
\includegraphics [angle=-90,scale=1]{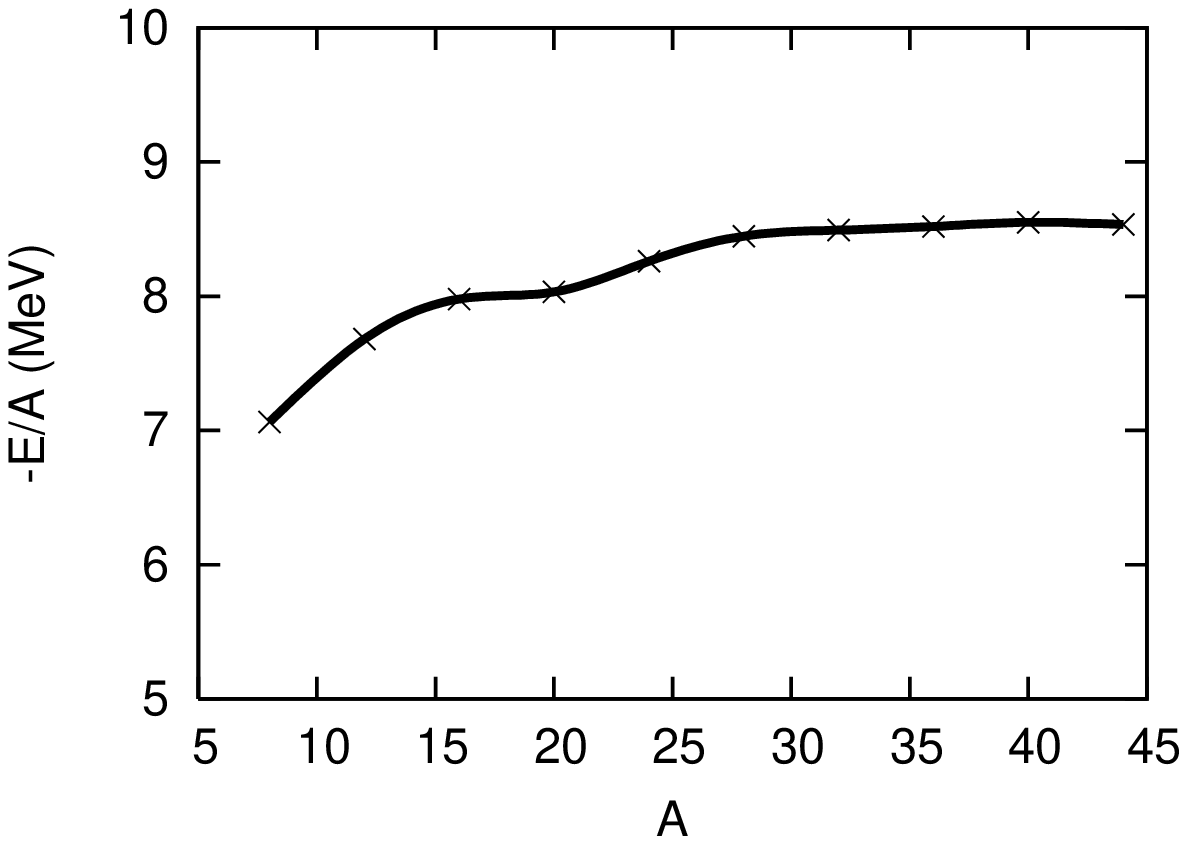}
\caption{Binding energy per nucleon for spin-isospin saturated nuclei 
for the SWCP model.}
\label{Fig4}
\end{center}
\end{figure}

We show in Fig.~\ref{Fig4} the results for $\be$ obtained in 
the SWCP model with the use of the density dependent 
Pauli potential, Eq. (\ref{VPden}). The statistical uncertainty of these calculated values is at most 5 \%. 
%%%%%%%%%%%%%%%%%%%%%%%%%%%%%%%%%%%%%%%%%%%%%%%%%%%%%%%%%%%%%%%%%
%%%%%%%%%%%%%%%%%%%%%%%%%%%%%%%%%%%%%%%%%%%%%%%%%%%%%%%%%%%%%%%%%

\begin{figure}[hbtp]
\begin{center}
\includegraphics [scale=2] {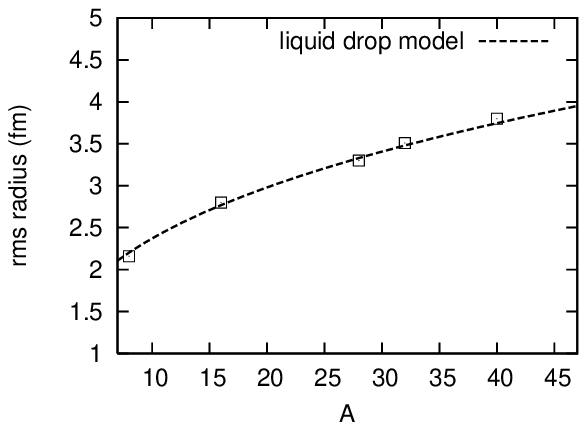}
\caption{Matter root mean square radius calculated as a function of mass number A for several nuclei using the SWCP model. See details in the text. Liquid drop model calculation is shown for comparison.} 
\label{Fig21}
\end{center}
\end{figure}

In Fig.~\ref{Fig21}  we show the matter rms radius (boxes) as a function of mass number, A, for a set of nuclei as calculated with the density dependent fit in Eq. (\ref{VPden}). We can see that they approximately follow the liquid drop model results (dashed line) given by the $R=r_0 A^{1/3}$ with $r_0\approx 1.1 fm$.

\begin{figure}[hbtp]
\begin{center}
\includegraphics [scale=1.5] {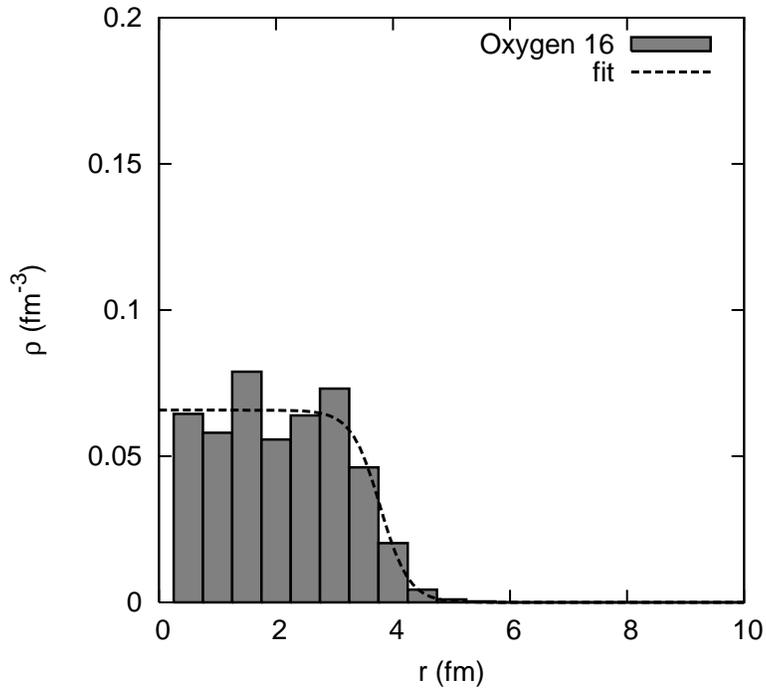}
\caption{$^{16}$O density distribution histogram as given by the SWCP model. Dashed line shows the Woods-Saxon fit for this nucleus. See details in the text. } 
\label{Fig22}
\end{center}
\end{figure}

\begin{figure}[hbtp]
\begin{center}
\includegraphics [scale=1.5] {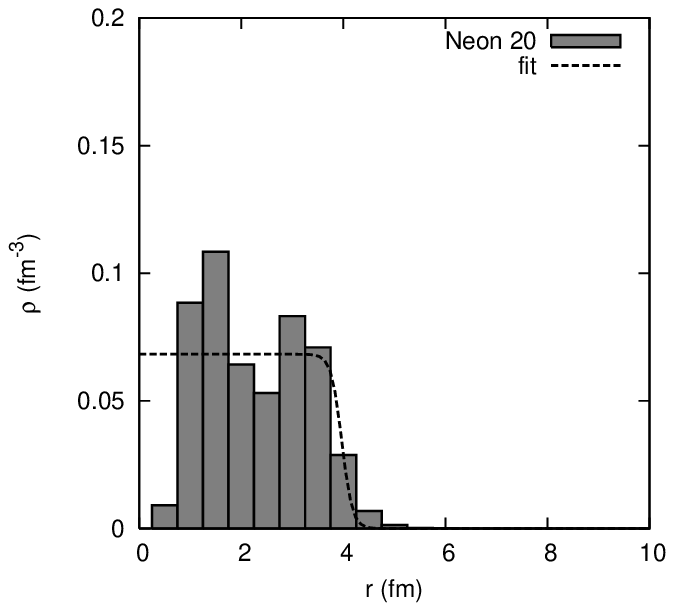}
\caption{Same as Fig. \ref{Fig22} for $^{20}$Ne.} 
\label{Fig23}
\end{center}
\end{figure}

In Figs.~\ref{Fig22} and ~\ref{Fig23}  we show the density distribution calculated for an $^{16}$O nucleus and a $^{20}$Ne nucleus, respectively. We can see that most of the mass is located in the nuclear central region. The decrease of particle population at small distances is due to the strong repulsive core of the interaction used. We have fit this distribution to a typical Woods-Saxon form $\rho(r)=\frac{\beta}{1+e^{(r-\alpha)/\gamma}}$  obtaining a set of values for $^{16}$O, $\beta=0.066\pm 0.002 fm^{-3}$, $\alpha=4.0\pm 0.5 fm$ and $\gamma=0.41\pm 0.45 fm$ and for $^{20}$Ne we get $\beta=0.068\pm 0.007 fm^{-3}$, $\alpha=3.97\pm 0.20 fm$ and $\gamma=0.1\pm 0.5 fm$ which are in reasonable agreement with those of mean field calculations~\cite{QHD}. Note that the value $\alpha$ indicates where the central density remains approximately constant, while the matter rms radius gets contribution also from the more extended spatial region of nucleons in the nucleus. It is worth noting that the values of the parameter $a$ and $b$ are mainly responsible for reproducing the rms radii in this approach.

%%%%%%%%%%%%%%%%%%%%%%%%%%%%%%%%%%%%%%%%%%%%%%%%%%%%%%%%%%%%%%%%%
%%%%%%%%%%%%%%%%%%%%%%%%%%%%%%%%%%%%%%%%%%%%%%%%%%%%%%%%%%%%%%%%%

%%%%%%%%%%%%%%%%%%%%%%%%%%%%%%%%%%%%%%%%%%%%%%%%%%%%%%%%%%%%%%%%%
%%%%%%%%%%%%%%%%%%%%%%%%%%%%%%%%%%%%%%%%%%%%%%%%%%%%%%%%%%%%%%%%%
%%%%%%%%%%%%%%%%%%%%%%%%%%%%%%%%%%%%%%%%%%%%%%%%%%%%%%%%%%%%%%%%%
%%%%%%%%%%%%%%%%%%%%%%%%%%%%%%%%%%%%%%%%%%%%%%%%%%%%%%%%%%%%%%%%%

\begin{figure}[hbtp]
\begin{center}
\includegraphics [angle=-90,scale=1] {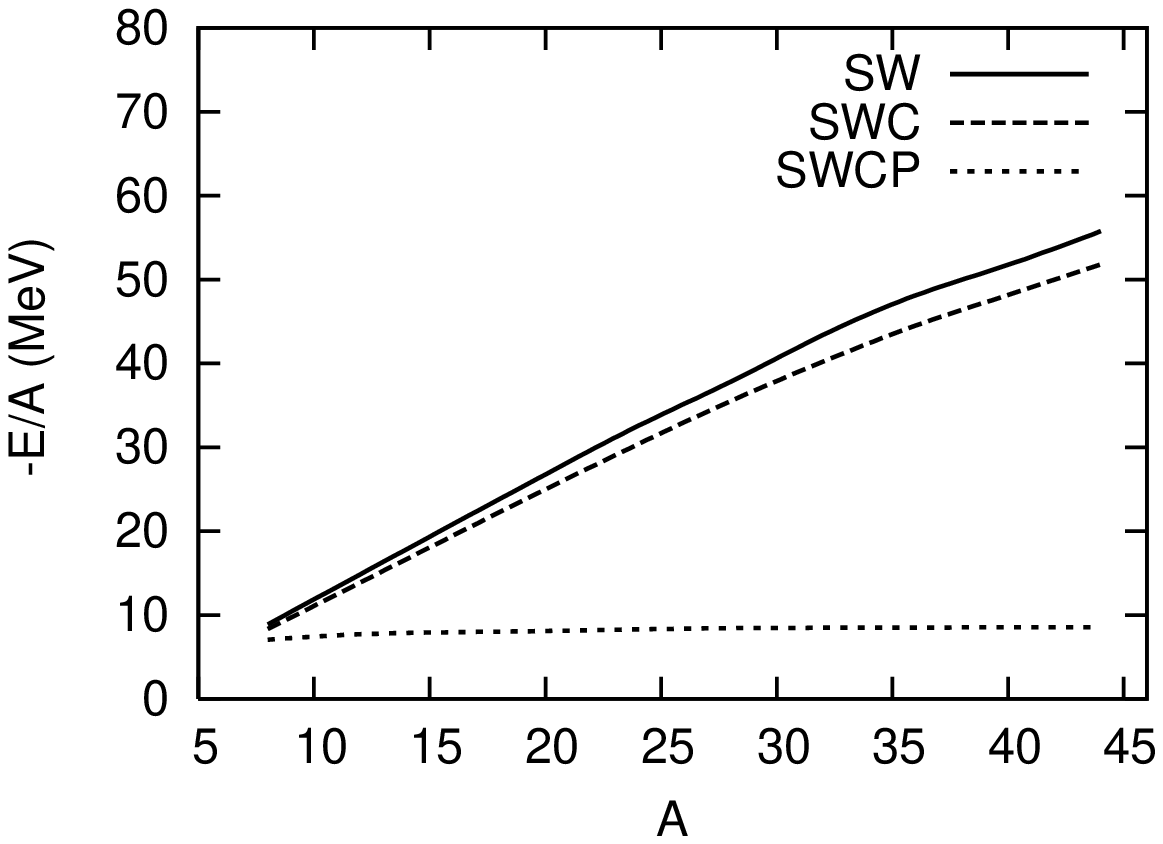}
\caption{Binding energy per nucleon ($\be$) for different NN 
interaction models.}
\label{Fig5}
\end{center}
\end{figure}

Next, we present in Fig.~\ref{Fig5} the comparison for the binding energy per nucleon obtained with the different models described in Sect.~\ref{SWmodel}.
%%%%%%%%%%%%%%%%%%%%%%%%%%%%%%%%%%%%%%%%%%%%%%%%%%%%%%%%%%%%%%%%%
%%%%%%%%%%%%%%%%%%%%%%%%%%%%%%%%%%%%%%%%%%%%%%%%%%%%%%%%%%%%%%%%%
The solid (long-dashed) [short-dashed] line shows the SW (SWC) [SWCP] model. 
Both, the SW and SWC models, cannot achieve the saturation 
for $\be$. The Coulomb potential can negligibly help to achieve 
$\be$ saturation as expected, although it is important 
for some properties of finite nuclei~\cite{QHD}.

\begin{figure}[hbtp]
\begin{center}
\includegraphics [angle=-90,scale=1] {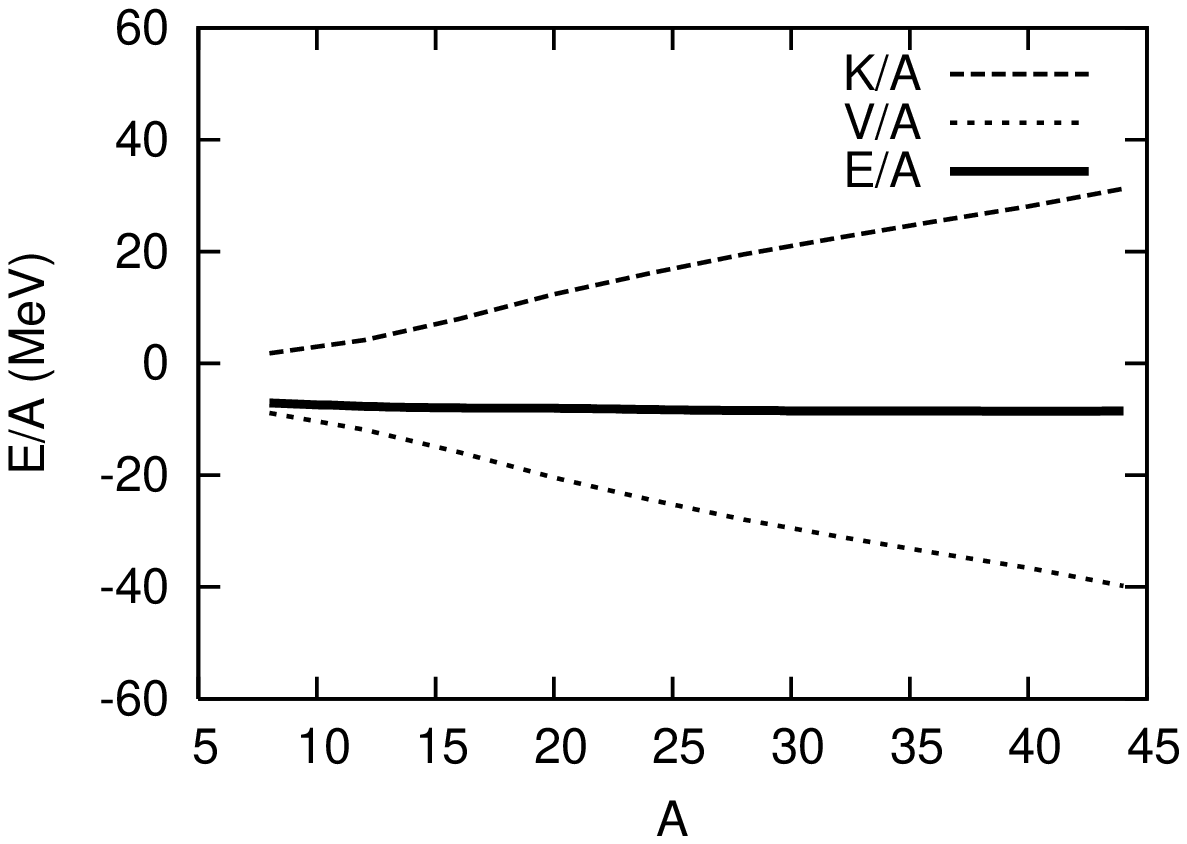}
\caption{Kinetic ($K/A$) and potential ($V/A$) energy per nucleon 
contributions to $E/A$.}
\label{Fig6}
\end{center}
\end{figure}

In order to understand how the energy saturation is obtained
in the SWCP model, we draw separately in Fig.~\ref{Fig6} the contribution
of the kinetic energy per nucleon $K/A$ (dashed line), and potential 
energy per nucleon $V/A$ (short-dashed line)
together with $E/A$ (solid line).
%%%%%%%%%%%%%%%%%%%%%%%%%%%%%%%%%%%%%%%%%%%%%%%%%%%%%%%%%%%%%%%%%
%%%%%%%%%%%%%%%%%%%%%%%%%%%%%%%%%%%%%%%%%%%%%%%%%%%%%%%%%%%%%%%%%
We see the kinetic energy per nucleon largely deviates 
from the classical value, $K/A=(3/2) k_B T$, (i.e. $K/A=1.5$ MeV 
in the present case with $T=1$ MeV) due 
to the contribution of the density dependent Pauli potential.
Thus, this shows that the kinetic and canonical momenta are completely 
different as the nucleon density increases~\cite{newpauli}. 
A balance between the large potential and 
kinetic energy contribution leads to $\be$ saturation. This cancellation arises 
as a consequence of the increasing kinetic contribution of Pauli correlations~\cite{future}.
This may be analogous to the mechanism of saturation in 
relativistic mean field models which involves the cancellation between 
the attractive scalar and the repulsive vector potentials~\cite{QHD}.

\begin{figure}[hbtp]
\begin{center}
\includegraphics [angle=-90,scale=1]{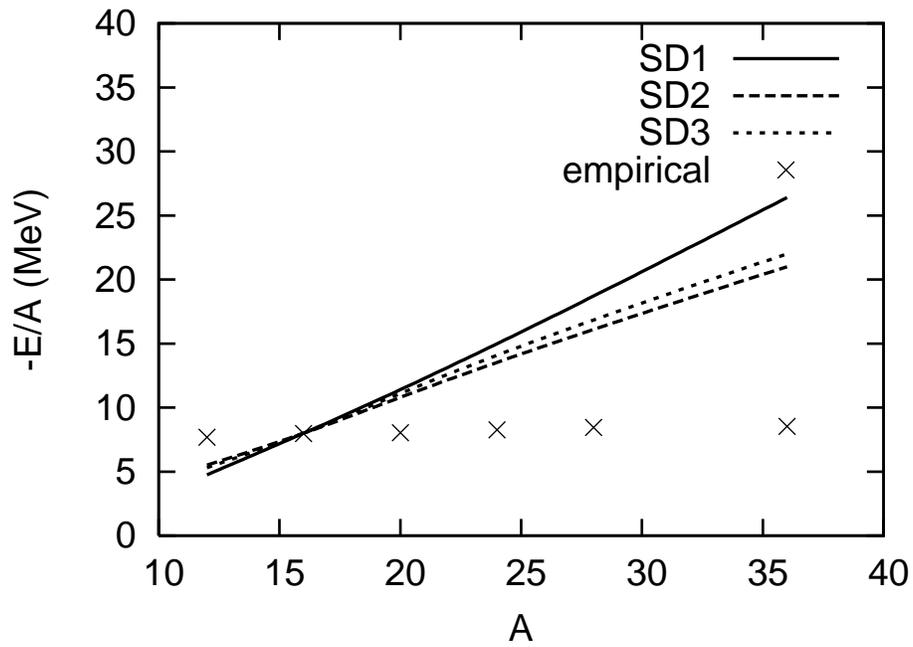}
\caption{Same as Fig.~\ref{Fig2} 
for the SD1, SD2 and SD3 models.}
\label{Fig7}
\end{center}
\end{figure}

Let us now study the effect of spin dependent potentials, 
using the SD1, SD2 and SD3 models described in 
Sect.~\ref{SDmodel}. In this case we will work
with the density independent Pauli potential described
in Sect.~\ref{SWmodel}, in order to see the effect 
of spin-dependent forces in the energy saturation. 
The results are shown in Fig.~\ref{Fig7}.
%%%%%%%%%%%%%%%%%%%%%%%%%%%%%%%%%%%%%%%%%%%%%%%%%%%%%%%%%%%%%%%%%
%%%%%%%%%%%%%%%%%%%%%%%%%%%%%%%%%%%%%%%%%%%%%%%%%%%%%%%%%%%%%%%%%
The solid (long-dashed) [short-dashed] line stands for 
the results of the SD1 (SD2) [SD3] model,  
together with the empirical data (crosses)~\cite{nucdata}.
We can see again the $\be$ saturation cannot be obtained
even using an improved spin dependent NN potential. 

\begin{figure}[hbtp]
\begin{center}
\includegraphics [angle=-90,scale=1]{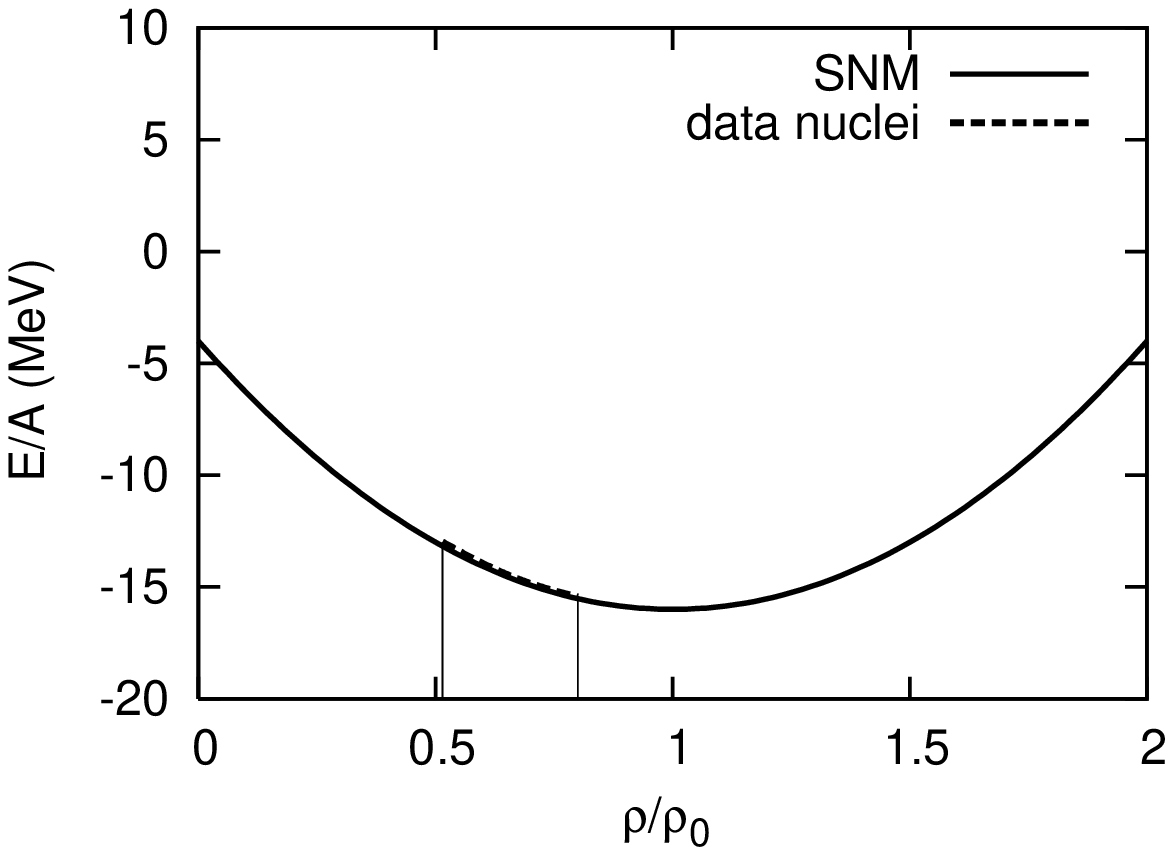}
\caption{Energy per particle in SNM at $T=0$ and range in densities corresponding to the Fermi momentum of nuclei considered in this work. The lower bound in the density interval corresponds to $\rho/\rho_0=0.51$ and the upper to $\rho/\rho_0=0.8$. } 
\label{Fig8}
\end{center}
\end{figure}

In Fig.~\ref{Fig8} we show energy per nucleon for the ideal "infinite" symmetric nuclear matter  (SNM) case at $T=0$ MeV (solid line). An incompressibility value of $K\approx 230$ MeV has been assumed and $\rho_0=0.165 fm^{-3}$.  Correspondingly the range in densities or Fermi momentum can be extracted as done in ref.\cite{moniz} as $\rho=\frac{2p_F^3}{3\pi^2}$ for the set of symmetric spin saturated nuclei $8 \le A \le 44$ (dashed line). The lower and upper bounds in the density interval correspond to $\rho/\rho_0=0.51$ and $\rho/\rho_0=0.8$ respectively.

%%%%%%%%%%%%%%%%%%%%%%%%%%%%%%%%%%%%%%%%%%%%%%%%%%%%%%%%%%%%%%%%%
%%%%%%%%%%%%%%%%%%%%%%%%%%%%%%%%%%%%%%%%%%%%%%%%%%%%%%%%%%%%%%%%%

\begin{figure}[hbtp]
\begin{center}
\includegraphics [angle=-90,scale=1]{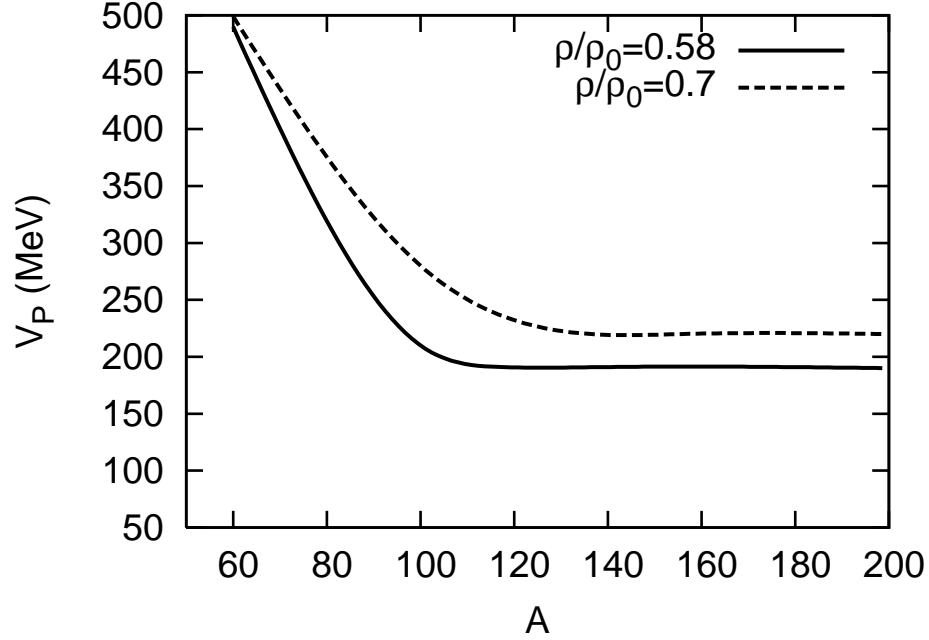}
\caption{Pauli potential strength as a function of the simulation system particle number for SNM for densities $\rho/\rho_0=0.58$ and $\rho/\rho_0=0.7$.}
\label{Fig9}
\end{center}
\end{figure}

\begin{figure}[hbtp]
\begin{center}
\includegraphics [angle=-90,scale=1]{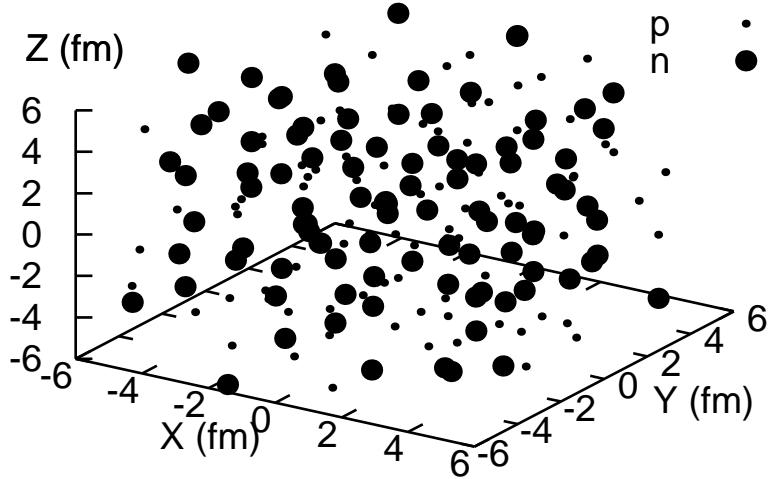}
\caption{Simulation box for a system of SNM consisting of $200$ particles at a density $\rho/\rho_0=0.7$ and $T=1$ MeV. Protons and neutrons are depicted with small and larger dots respectively.}
\label{Fig10}
\end{center}
\end{figure}

We now compare some results from heavier nuclear systems than the medium mass range nuclei considered in the previous sections in this work. In Fig.~\ref{Fig9} we show the Pauli potential strength, $V_P$ as a function of the particle number, $A$, for an idealized case of SNM,  through the simulation for densities $\rho/\rho_0=0.58$ (solid line) and $\rho/\rho_0=0.7$ (dashed line). We can see that finite size effects (particle number)  are important up to values about $A\approx 120$ where the Pauli potential strength stabilize to an approximately constant value. The larger values for small $A$ do not match those of finite nuclei since in this case nucleons are artificially constrained in a cube of length $L=(A/\rho)^{1/3}$. We use periodic boundary conditions as usual to minimize boundary effects near the walls of the simulation box. It is worth mentioning that as we consider heavier nuclei~\cite{future} the density dependence parametrized by the Fermi momentum must be replaced by the system size, $A$, since one can map out Pauli potential strengths in a clearly defined way. In Fig.~\ref{Fig10} we show a thermalized configuration of a simulation of a SNM plasma at $T=1$ MeV at a density $\rho/\rho_0=0.7$. In this homogeneous system of $A=200$ particles, protons are shown as small dots while neutrons correspond to larger dots.
 
%%%%%%%%%%%%%%%%%%%%%%%%%%%%%%%%%%%%%%%%%%%%%%%%%%%%%%%%%%%%%%%%%
%%%%%%%%%%%%%%%%%%%%%%%%%%%%%%%%%%%%%%%%%%%%%%%%%%%%%%%%%%%%%%%%%

%%%%%%%%%%%%%%%%%%%%%%%%%%%%%%%%%%%%%%%%%%%%%%%%%%%%%%%%%%%%%%%%%
%%%%%%%%%%%%%%%%%%%%%%%%%%%%%%%%%%%%%%%%%%%%%%%%%%%%%%%%%%%%%%%%%

\section{Summary and conclusions}
\label{summary}
Using many-body simulations with Monte Carlo techniques we have pursued the study of an effective Pauli potential in the nuclear binding energy and matter rms radius considering density dependence. The density dependent effective Pauli potential  partially simulates the effect of genuine fermionic correlations on semiclassical descriptions. For medium mass range nuclei with $ 8 \le A \le 44$ the nucleon number dependence into the effective Pauli potential can be replaced in terms of the Fermi momentum of the constituent nucleons. Even if spin dependent effects are explicitly included in the NN potential, an appropriate parametrization of the density dependence can still be derived in terms of the Fermi momentum. The density dependence in the effective Pauli potential is crucial to reproduce nuclear binding energy saturation, wich is achieved  by a balance of density dependent attractive and repulsive contributions in an analogous way to the relativistic mean field approach. Although for medium mass nuclei the density dependence can be described in terms of the Fermi momenta, the treatment of heavier nuclei requires to consider the density dependence in terms of the nucleon number, due to saturation of the Fermi momenta around 270 MeV/c. 

The use of spin-isospin dependent NN potentials 
turns out to give only a moderate 
improvement for the binding energy per nucleon,
but it is unable to achieve energy saturation
unless a density dependent Pauli potential is used. 
Our results show that, provided a different set of values 
for the spatial parameters, $a$ and $b$ or 
potential well strength, $V_0$, in the NN potential, 
a $p_F$ dependent Pauli potential strength
can always be parametrized to 
reproduce the empirical nuclear binding energies. 
The procedure presented in this work is robust.
Then, each set of nuclei modelled by a given 
NN potential and calibrated by the empirical binding energies, 
can be tested by studying further properties 
of nuclei like, for example, when introducing some hyperonic content.

We have shown that, in the density range close to the saturation of nuclear binding energies, and for an idealized SNM system the Pauli potential strength depends on the size of the system and stabilizes to an approximately constant value around $A\approx 120$.
The matter rms radii of medium mass range nuclei can be satisfactorily reproduced by tuning the spatial parameters of the NN interaction to a reference nucleus (in this work taken to be  $^{16}$O). A close agreement with the typical sizes of nuclei expected from a liquid drop model is obtained.

To summarize, the present approach may provide a practical efficient way to model a set of spin-isospin symmetric nuclei calibrated by the empirical binding energies and matter rms radii. It also can be regarded as a preliminary step for studying further properties of symmetric many-nucleon 
systems in semiclassical simulations. It is necessary to provide further insight into semiclassical effective potential models to study wether asymmetric systems can be treated in a similar fashion. %%%%%%%%%%%%%%%%%%%%%%%%%%%%%%%%%%%%%%%%%%%%%%%%%%%%%%%%%%%%%%%%

\vspace{2ex}

\noindent{\bf Acknowledgments}\\
We would like to thank Tomoyuki Maruyama and A.W. Thomas 
for helpful discussions.
M.A.P.G. would like to dedicate this work to the memory 
of J.M.L.G. This work has been partially funded by the Spanish Ministry 
of Education and Science projects DGI-FIS2006-05319, SAB2005-0059 and FPA2007-65748,
and by Junta de Castilla y Le\'on under contracts SA-104104 and SA016A07. This work has been supported by the Spanish Consolider-Ingenio 2010 Programme CPAN (CSD2007-00042).

%%%%%%%%%%%%%%%%%%%%%%%%%%%%%%%%%%%%%%%%%%%%%%%%%%%%%%%%%%%%%%%%
% Bibliography. 
%%%%%%%%%%%%%%%%%%%%%%%%%%%%%%%%%%%%%%%%%%%%%%%%%%%%%%%%%%%%%%%%

%%%%%%%%%%%%%%%%%%%%%%%%%%%%%%%%%%%%%%%%%%%%%%%%%%%%%%%%%%%%%%%%
\end{document}